# Effects of non-standard interaction on microscopic black holes from ultra-high energy neutrinos

Ashutosh Kumar Alok[a], Trisha Sarkar[b], Shweta Yadav[c]

Indian Institute of Technology Jodhpur, Jodhpur 342037, India

**Abstract** If the universe has more than 4-dimensions, the TeV scale gravity theories predict formation of microscopic black holes due to interaction of ultra high energy neutrinos coming from some extragalactic origin with the nucleons present in the Earth's atmosphere. The decay of these black holes can generate high multiplicity events which can be detected through neutrino telescopes. Ultra high energy neutrinos can also produce events without the formation of black holes which can be distinguished from the black hole events depending on their topological structure. In this work we study the effects of non-standard interaction on the production of these shower events. We find that new physics has inconsequential impact on the number of events produced through the generation of black holes. For events produced without the formation of black holes, new physics can only provide a marginal deviation. Therefore a large enhancement in the number of shower events over the standard model prediction can provide unambiguous signatures of TeV scale gravity in the form of microscopic black hole production.

## 1 Introduction

One of the primary goals of probing physics beyond the standard model (SM) is to resolve the problem of hierarchy at the electroweak scale, i.e., the existence of large disparity between the weak scale and gravity. One of the possible solutions of the hierarchy problem is to consider the Planck scale to be as small as TeV scale which also satisfies the naturalness condition. This can lead to the presence of large extra dimensions (LED) in addition to our three spatial dimensions and one temporal dimension. From this perspective the SM is considered to be confined to a 3-brane in a hyperspace of higher dimensions, $D = 4 + n$, $n$ being the number of extra dimensions. Such contemplation elucidates the relatively weaker strength of gravity as it can propagate along all the additional spatial dimensions unlike the SM gauge fields. Since gravity is considered to be unified with the other fundamental interactions at the Planck scale, the effects of quantum gravity can be studied at the particle accelerators such as the LHC.

One of the most striking consequences of the TeV-scale gravity is the formation of microscopic black holes (BHs) created in ultra high energy (UHE) particle collisions [1]. When the center of mass (CM) energy of the two UHE particle collisions exceeds the Planck energy scale, there is a possibility of creation of a microscopic BH. If the Planck energy is scaled down to the TeV range, it is possible to generate such tiny BHs at the LHC. However, since the current maximum energy reach of the artificial particle colliders is limited, UHE cosmic neutrinos are exceptional sources to probe microscopic BHs as they can achieve much greater energy.

Earth is bombarded with UHE cosmic ray particles every year with energies above $10^8$ GeV. UHE neutrinos produced from different sources of cosmic accelerators collide with the nucleons in the atmosphere or inside the Earth matter providing CM energy exceeding $\sim$ 100 TeV. At such high energies, neutrinos experience deep inelastic scattering with the nucleons which can produce microscopic BHs in the Earth atmosphere in the presence of LEDs. These BHs have very small lifetime and they undergo evaporation almost instantaneously into the SM or BSM particles by Hawking radiation [2]. Such phenomena is expected to give rise to high multiplicity events [3] resulting in large air showers which are detected by particle detectors [1] such as IceCube [4] or ANITA [5] which can detect neutrinos having energy above PeV range. The shower characteristics corresponding to the BH events are different from the SM shower events based on the event topologies and the final state particles [6].

[a] e-mail: akalok@iitj.ac.in
[b] e-mail: sarkar.2@iitj.ac.in (corresponding author)
[c] e-mail: yadav.21@iitj.ac.in

In the Ref. [7], it was shown that UHE neutrinos collide with the nuclei present in the the Earth or the atmosphere and produce electrically charged or neutral leptons $l$ and a hadronic shower $X$ in the process. In presence of LEDs, the production of these leptons and hadrons are accompanied by an intermediate resonance state of microscopic BH which decays almost immediately and give rise to a large number of decay products. The events generated in the process every year are detectable by the large neutrino telescope such as the IceCube detector. Further, the number of events produced via the BH channel was calculated in presence of different number of LEDs. It was shown that the number of BH events are usually much higher than those produced in absence of BH production. The results differ noticeably for two different types of incoming neutrino flux.

The results in [7] were obtained assuming that the interactions are governed by the SM physics. In this work we include the effects of new physics neutrino interactions manifested in the form of non standard interaction (NSI) on these two types of events to observe whether the new physics can alter some of the results obtained in the Ref. [7]. NSI can be incorporated in the effective Lagrangian in terms of higher dimensional Lorentz structure ($d > 4$) invariant under SM gauge group $(SU(3)_c \otimes U(1)_{EM})$ while in the SM, only the operators with dimension $d \leq 4$ are allowed. In this work, we restrict to dimension-6 operators. Since the charged current (CC) NSI is strictly constrained, the effect of NSI is only expected to affect the neutral current (NC) interaction [8]. The bounds on NSI parameters have been obtained by global analysis of the data from different oscillation and scattering experiments like COHERENT [9–12]. Bounds from deep inelastic scattering are also obtained from the analysis of CHARM data [13–15]. It would be interesting to see whether the current bounds on NSI parameters allow observable effects on the events generated though the BH production within the framework of TeV scale gravity theory.

The plan of the work is as follows. In Sect. 2, we briefly describe the NSI and its effect on neutrino nucleon scattering. We also describe the phenomena of microscopic BH formation and decay. Further, we provide an outline of the formation of shower events. In Sect. 3 we demonstrate our results whereas we conclude in Sect. 4.

## 2 Formalism

In this section we provide the theoretical framework of the analysis. We start with a brief description of NSI and its implication on neutrino interactions in Sect. 2.1. Then in the Sect. 2.2, we briefly describe the microscopic BH phenomenology in presence of LEDs. In Sect. 2.3, we present a layout of the generation of the shower events relevant to this work.

### 2.1 Non standard interaction

The concept of NSI emerges from various BSM scenarios and neutrino mass models, see for e.g., [16–19]. Considering SM to be a lower energy approximation of some full theory valid at higher energy scale, the effective Lagrangian can be represented in terms of higher dimensional operators as

$$\mathscr{L}_{eff}^{(4)} = \mathscr{L}_{SM}^{(4)} + \frac{1}{\Lambda} C_i^{(5)} \mathscr{O}_i^{(5)} + \frac{1}{\Lambda^2} C_i^{(6)} \mathscr{O}_i^{(6)} + \cdots \quad (1)$$

Here $\Lambda$ is the scale of new physics and $C_i$'s are the Wilson coefficients corresponding to different four-fermi operators $\mathscr{O}_i$. These Wilson coefficients incorporate the characteristics of short distance physics. In our analysis we consider dimension-6 four fermion vector operators. In the lower energy limit, in a model independent framework, such operators for CC and NC neutrino interactions are represented as [17,20,21]

$$\begin{aligned}\mathscr{L}_{CC-NSI} &= 2\sqrt{2} G_F \varepsilon_{\alpha\beta}^{ff',X} (\bar{\nu}_\alpha \gamma^\mu P_L l_\beta)(\bar{f'} \gamma_\mu P_X f), \\ \mathscr{L}_{NC-NSI} &= 2\sqrt{2} G_F \varepsilon_{\alpha\beta}^{f,X} (\bar{\nu}_\alpha \gamma^\mu P_L \nu_\beta)(\bar{f} \gamma_\mu P_X f). \end{aligned} \quad (2)$$

Here $\varepsilon_{\alpha\beta}^{ff',X}$ and $\varepsilon_{\alpha\beta}^{f}$ ($\alpha, \beta = e, \mu, \tau$) with $X \in \{L, R\}$ are the dimensionless NSI parameters which determine the strength of new physics. Further, $P_{L,R} = (1 \mp \gamma^5)/2$ are the left and right handed chirality operators. For CC-NSI, $f \neq f'$, $f, f' = u, d$ whereas for NC-NSI, $f = f'$, $f = e, u, d$. Since CC-NSI is strictly constrained at least by an order of magnitude compared to NC-NSI due to the bounds coming mainly from the Fermi constant, CKM unitarity, pion decay and the kinematic measurements of the masses of the gauge bosons $M_Z$ and $M_W$, we consider NSI effects only in NC-NSI sector. In the limit $\varepsilon_{\alpha\beta} \to 0$, SM results are reimposed. Vector ($V$) and axial-vector ($A$) NSI parameters are defined as, $\varepsilon_{\alpha\beta}^{f,V(A)} = \varepsilon_{\alpha\beta}^{f,L} \pm \varepsilon_{\alpha\beta}^{f,R}$. If $\varepsilon_{\alpha\beta}^{f} \neq 0$ for $\alpha \neq \beta$, then it implies lepton flavour violation (LFV), while $\varepsilon_{\alpha\beta}^{f} \neq 0$ indicates lepton flavour universality violation (LFUV). For detailed review on NSI, see for e.g., [19,21].

The bounds on NSI parameters are usually obtained from the global analysis of neutrino oscillation data [22]. However, the oscillation data cannot distinguish between the two degenerate mixing angle solutions: LMA-Light ($\theta_{12} \approx 34°$) and LMA-Dark ($45° < \theta_{12} < 90°$). Particularly, the LMA-Dark solution is responsible for the large values of the NSI parameters, $\mathscr{O}(\varepsilon) \sim 1$. The presence of such degeneracy foils the accurate determination of the $CP$ violating phase and the sign of the larger mass-square difference ($\Delta m_{31}^2$) which is one of the main objectives of the neutrino oscillation experiments [9,14]. Therefore the bounds on the NSI parameters are also obtained from the various scattering experiments. The combined bounds attained from both the oscillatory as well as scattering experiments lift the LMA-Light and LMA-Dark degeneracy [11,14].

NSI can be generated in several new physics models below the electroweak scale, specially the models which are renormalizable having an additional $U(1)'_X$ symmetry with a $Z'$ gauge boson as a mediator. There is possibility for the mediator to be light ($\mathcal{O}(10\text{MeV}) \leq M_{Z'} \leq \mathcal{O}(1\text{GeV})$) as well as heavy ($\mathcal{O}(1\text{GeV}) \leq M_{Z'} \leq \mathcal{O}(1\text{TeV})$) [9]. In the case of light mediator, NSI bounds can be obtained from the coherent neutrino-nucleon scattering (CE$\nu$NS) in COHERENT experiment. For heavy $Z'$, the bounds from the experiments such as CHARM and NuTeV are also taken into account in addition to the COHERENT result [9].

NSI affects neutrino nucleon interaction for coherent elastic as well as deep inelastic scattering. In our work, we follow Ref. [13] for the analysis of neutrino-nucleon deep inelastic scattering for large neutrino energy ($E_\nu$) and momentum exchange ($Q$). The full basis of dimension-6 effective operators below the EW scale are represented as

$$\mathcal{O}^{(6)}_{1,f} = (\bar{\nu}_\alpha \gamma^\mu P_L \nu_\beta)(\bar{f}\gamma_\mu f),$$
$$\mathcal{O}^{(6)}_{2,f} = (\bar{\nu}_\alpha \gamma^\mu P_L \nu_\beta)(\bar{f}\gamma_\mu \gamma^5 f). \quad (3)$$

The linear combination of $\mathcal{O}^{(6)}_{1,f}$ and $\mathcal{O}^{(2)}_{2,f}$ produce the operator of $\mathcal{L}_{NC-NSI}$ in Eq. (2) i.e. $(\mathcal{O}^{(6)}_{1,f} - \mathcal{O}^{(6)}_{2,f})$ for $X = L$ and $(\mathcal{O}^{(6)}_{1,f} + \mathcal{O}^{(6)}_{2,f})$ for $X = R$. The corresponding Wilson coefficients contain the contributions from both SM and NSI as, $C^{(6)}_{i,f} = C^{(6)}_{i,f}|_{SM} + C^{(6)}_{i,f}|_{NSI}$ where $i = 1, 2$. For the NC transition $\nu_\alpha \to \nu_\beta$, the SM contributions are given by

$$C^{(6)}_{1,u}|_{SM} = -C^{(6)}_{2,u}|_{SM} + \frac{4\sqrt{2}}{3}G_F s_w^2 \delta_{\alpha\beta},$$
$$C^{(6)}_{1,d}|_{SM} = -C^{(6)}_{2,d}|_{SM} - \frac{2\sqrt{2}}{3}G_F s_w^2 \delta_{\alpha\beta} = C^{(6)}_{1,s}|_{SM},$$
$$C^{(6)}_{2,u}|_{SM} = \frac{G_F}{\sqrt{2}}\delta_{\alpha\beta}, \quad C^{(6)}_{2,(d,s)}|_{SM} = -C^{(6)}_{2,u}|_{SM}, \quad (4)$$

while the NSI contribution is expressed as

$$C^{(6)}_{1(2),f}|_{NSI} = \frac{G_F}{\sqrt{2}}\varepsilon^{f,V(A)}_{\alpha\beta}. \quad (5)$$

In Eq. (5), $f = u, d, s$ and $V, A$ denote the vector and axial vector part, respectively. Also, $s_w = \sin\theta_w$, where $\theta_w \sim 28.13°$ is the Weinberg angle. The expression for differential scattering cross section is given as [13]

$$\frac{d^2\sigma^{CC}}{dxdy} = \frac{G_F^2 m_N E_\nu}{\pi}$$
$$\times \left[ xu(x) + xd(x) + 2s(x) \right.$$
$$\left. + (1-y)^2\{\bar{u}(x) + \bar{d}(x)\} \right]$$

$$\frac{d^2\sigma^{NC}}{dxdy} = \frac{m_N E_\nu}{\pi}$$
$$\times \left[ (C_{L,u}^2 + C_{L,d}^2) \sum_{f=u,d}\{xf(x) \right.$$
$$+ x\bar{f}(x)(1-y)^2\} + (C_{R,u}^2 + C_{R,d}^2)$$
$$\times \sum_{f=u,d}\{xf(x)(1-y)^2 + x\bar{f}(x)\}$$
$$+ 2\{C_{L,s}^2(xs(x) + \bar{s}(x)(1-y)^2)$$
$$\left. + C_{R,s}^2(xs(x)(1-y)^2 + \bar{s}(x))\} \right], \quad (6)$$

Equation (6) implies that NSI only affects NC interaction. Although flavour conserving processes are allowed in both SM and NSI, they have different values of Wilson coefficients. FCNC processes are allowed in NSI, but suppressed in the SM.

### 2.2 Microscopic black hole phenomenology

As discussed earlier in Sect. 1, one of the possible solutions to the hierarchy problem is acquired by introducing the concept of brane scenario in which the SM matter and gauge fields can only propagate in a 3-brane, confined in a higher dimensional space. The extra spatial dimensions ($n$) are compact and large as compared to the weak scale [23]. Thus the fundamental Planck scale at 4D sub-manifold ($M_4^{Pl} \sim 10^{19}$ GeV) is reduced to the effective Planck scale ($M_{Pl,4+n}$) as [23]

$$M_{Pl,4}^2 \sim R^n M_{Pl,4+n}^{2+n}, \quad (7)$$

where $R$ is the length scale of the extra dimensions. Gravity is mediated in the extra bulk dimensions via $(4+n)$ dimensional graviton whose coupling is constrained from the effective Planck scale.

One of the most important tool to study the nature of strong gravitational field is the BH as it is assumed that gravity is strongest at the singularity existing at the center of the BH.

According to the Hoop Conjecture [24], when the impact parameter $b$ of the collision between two partons $i$ and $j$ is much smaller than the Schwarzschild radius $r_s$ associated with the CM of the scattering process, there is a possibility of BH formation if the particles pass within the event horizon before they come in causal contact [25]. As the BH formed in this process is associated with the energy exceeding the Planck scale, semi-classical and thermodynamics description is required [25,26]. A microscopic BH created in high energy particle collisions having radius smaller than the length scale of the extra spatial dimensions $R$ is well approximated by the $(4 + n)$ dimensional Schwarzschild BH solution [27]. The Schwarzschild radius of a $(4 + n)$ dimensional microscopic BH is expressed as [25,27,28]

$$r_{S(4+n)} = \frac{1}{M_{Pl(4+n)}}\left(\frac{M_{BH}}{M_{Pl(4+n)}}\right)^{\frac{1}{1+n}}$$

$$\times \left[\frac{2^n \pi^{(n-3)/2} \Gamma(\frac{n+3}{2})}{2+n}\right]^{1/1+n}. \tag{8}$$

Here $M_{BH}$ is the mass of the BH which is assumed to be $\sqrt{2m_N E_\nu x}$ [1,7]. Since the formation of a microscopic BH is a classical non-perturbative process, it is not associated with any short distance physics. Therefore the cross section of such process is represented by geometric cross section given by

$$\sigma_{ij \to BH} \approx \pi r_S^2. \tag{9}$$

A microscopic BH is considered to be an intermediate resonance state having the Hawking temperature, $T_H = (1+n)/4\pi r_S$ [27,28]. The decay of a BH is initiated by shedding the gauge quantum numbers obtained from the initial parton pairs which occurs via the emission of classical gauge radiation and gravitational radiation followed by the spin down phase in which it sheds the angular momenta. This Schwarzschild BH decays by Hawking radiation and emits a large number of SM particles mainly into the brane rather than the LEDs [25,26,29].

UHE cosmic neutrinos can produce microscopic BH in the atmosphere undergoing collisions with the nucleons. The corresponding scattering cross-section is given by [1,30]

$$\sigma(\nu N \to BH) = \sum_i \int_a^1 dx f_i(x) \times \pi r_S^2 = \sigma^{BH}, \tag{10}$$

where $i$ represents the partons and $a = (M_{min}^{BH})^2/2m_N E_\nu$. $M_{min}^{BH}$ is the minimum mass of the possible BH to be formed which is considered to be equal to $M_{Pl,4+n}$ TeV.

The basic concept is that the strength of gravity is increased and becomes equal to the other fundamental forces at the scale of quantum gravity. The number of extra dimensions ($n$) and the modified Planck scale ($E_{Pl,4+n}$) are constrained by many cosmological and astrophysical phenomena [31]. The case with only one extra dimension is ruled out as for $E_{Pl,4+n} \sim 1$ TeV, $R \sim 10^{13}$ cm from Eq. (7) which implies that gravity is modified over astronomical distances. For $n = 2$, $R \sim 1$ mm which is the lowest length scale over which gravity has been measured. In case of $n = 2$, the lower limit on $M_{Pl,4+n}$ was found to be exceeding 30 TeV [31–33] which is unreachable by the current accelerator experiments. For $n = 3$, $M_{Pl,4+n} \sim 2$ TeV [32]. For $n \geq 4$, the constraints come from cosmic ray and collider experiments [34] which shows $M_{Pl,4+n} \sim 1 - 2$ TeV. The analysis on the bound of the size of the LEDs have been performed in several works in the neutrino oscillation scenario considering the mixing between the three active neutrino flavours and three sterile neutrinos in context of several neutrino oscillation experiments [35–39].

### 2.3 Shower events

In neutrino detectors, the detected shower events are generated by the two processes, (1) $\nu N \to lX$ and (2) $\nu N \to BH \to lX$ which are shown in Fig. 1. Here $X$ represents hadrons while $l$ stands for the leptonic counterpart which can be both charged as well as neutral. The events generated from the evaporation of the intermediate BH state is higher in multiplicity as compared to without BH. Depending on the event topology observed in the detector, three different kinds of events are observed: (a) Shower like events which are produced by both CC and NC interactions of neutrinos, and the energy is accumulated in a nearly spherical structure, (b) track like events which are produced by muons from CC interaction and (c) double bang events generated by tauons coming from CC interaction of the neutrinos which appear as two distinct shower events [6,40]. As in this work we are interested in investigating NSI effects on number of events, only shower like events are relevant where NC interaction is involved along with CC interaction:

(1) Number of shower events produced in absence of BH formation via the process $\nu N \to lX$ is given by $N_{sh} = N_{\nu_e} + N_{\nu_\mu} + N_{\nu_\tau}$, where $N_{\nu_i}$ ($i = e, \mu, \tau$) is the number of shower events generated by the incoming neutrino of $i$-th

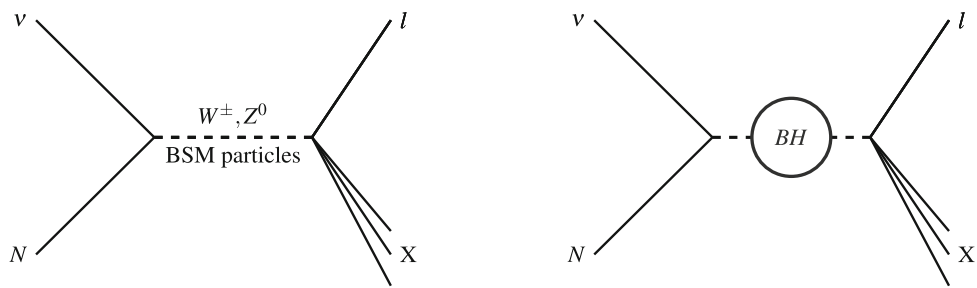

**Fig. 1** Feynman diagrams for neutrino nucleon scattering process. Left: in absence of BH production $\nu N \to lX$. In SM the process is mediated by $W^\pm$ or $Z^0$, depending on CC or NC interaction, while in presence of NSI, the interaction is mediated by any beyond standard model (BSM) particle; Right: in presence of BH production $\nu N \to BH \to lX$. Here $l$ is the charged or neutral lepton and $X$ is the shower counterpart

flavour, represented as

$$N_{\nu_i} = 2\pi AT \int \cos\theta_z \int dE_{\nu_i} \frac{d\phi_{\nu_i}}{dE_{\nu_i}} P_{surv}^i$$
$$\times \left( \int_{y_{i,CC}^{min}}^{y_{i,CC}^{max}} dy \frac{1}{\sigma^{CC}(E_{\nu_i})} \frac{d\sigma^{CC}(E_{\nu_i})}{dy} P_{int,i}^{CC} \right.$$
$$\left. + \int_{y_{i,NC}^{min}}^{y_{i,NC}^{max}} dy \frac{1}{\sigma^{NC}(E_{\nu_i})} \frac{d\sigma^{NC}(E_{\nu_i})}{dy} P_{int,i}^{NC} \right). \quad (11)$$

Here $y_{CC}^{min} = 0$, $y_{CC}^{max} = 1$, $y_{NC}^{min} = E_{sh}^{thr}$ and $y_{NC}^{max} = 1$. Further, $E_{sh}^{thr}$ is the threshold energy for the shower detection of the neutrino telescope, $A$ is the area of the detector and $T$ is the period of time over which the shower events are detected. $P_{int}$ is the probability of neutrino interaction in a detector of length scale $L$, which is given by

$$P_{int,i}^{CC,NC} = 1 - \exp\left(-N_A L \sigma^{CC,NC}(E_{\nu_i})\right), \quad (12)$$

where $N_A$ is the Avogadros number. $P_{surv}$ is the probability of a neutrino to reach the detector, represented as

$$P_{surv}^i = \exp\left[-X(\theta)N_A(\sigma^{CC}(E_{\nu_i}) + \sigma^{NC}(E_{\nu_i}))\right]. \quad (13)$$

Here $\theta$ is the zenith angle and $X(\theta)$ is the column density of the medium through which the neutrino passes to reach the detector.

(2) The shower events generated from the decay of the BH ($\nu N \to BH \to lX$) is given by $N^{BH} = N_{\nu_e}^{BH} + N_{\nu_\mu}^{BH} + N_{\nu_\tau}^{BH}$. Here $N_{\nu_i}^{BH}$ is the number of BH events generated by an incoming neutrino of flavour $i$ ($i = e, \mu, \tau$) and is given by

$$N_{\nu_i}^{BH} = 2\pi AT \int \cos\theta \int dE_{\nu_i} \frac{d\phi_{\nu_i}}{dE_{\nu_i}}(E_{\nu_i}) P_{surv}^{BH,i}$$
$$\times \int_{y_l}^{y_u} dy \frac{1}{\sigma^{BH}(E_{\nu_i})} \frac{d\sigma^{BH}}{dy}(E\nu_i) P_{int}^i, \quad (14)$$

where $\sigma_{BH}$ is the BH production cross section given by Eq. (10). Also, $y_l = 0$ and $y_u = 1$. $P_{int}^i$ is the probability of neutrino interaction to create the BH and is represented as

$$P_{int}^i = 1 - \exp[-N_A L \sigma^{BH}(E_{\nu_i})]. \quad (15)$$

$P_{surv}^{BH,i}$ is the survival probability of a neutrino to form a BH in the detector given as

$$P_{surv}^{BH,i} = \exp[-X(\theta)N_A(\sigma^{CC}(E_{\nu_i}) + \sigma^{NC}(E_{\nu_i}) + \sigma^{BH}(E_{\nu_i}))]. \quad (16)$$

In our work we have neglected the zenith angle dependence of the column density $X$. As discussed in Sect. 2.2, $\sigma^{BH}$ is independent of any short distance physics. Therefore the NSI effects on the shower events generated from the decay of a BH is incorporated through $\sigma^{NC}$ which is obtained from Eqs. (4), (5) and (6).

In Eqs. (11) and (14), $d\phi_{\nu_i}/dE_{\nu_i}$ is the incoming neutrino flux. The highest energy neutrinos ($\sim$ EeV) are predicted to be generated in the GZK process ($p\gamma \to n\pi^+$, $\pi^+ \to \mu^+\nu_\mu$, $\mu^+ \to e^+\nu_e\bar{\nu}_\mu$; $pp \to \pi^+ \to \mu^+\nu_\mu$) in gamma ray burst (GRB) and active galactic nucleus (AGN) [41–43]. The upper limit of such neutrino flux is restricted by Waxman-Bahcall flux (WB), $E_\nu^2 d\phi_\nu/dE_\nu < 2 \times 10^{-8}$ cm$^{-2}$s$^{-1}$sr$^{-1}$ [41]. The flux is similar for all three flavours of neutrinos, $\nu_e$, $\nu_\mu$, $\nu_\tau$ in the neutrino oscillation framework for a distant source [44].

## 3 Result and discussion

IceCube detector at south pole was able to detect high energy astrophysical neutrinos and identify its source [45–47].

The prospect of detecting microscopic BH events as well as ordinary shower events is considerably higher in such a large scale neutrino telescope due to its high sensitivity and detection technique. Here we study the impact of NSI on the number of BH and ordinary shower events in the context of IceCube experiment. The key ingredient in calculating the number of shower events lies in the scattering cross section. In our analysis, considering the lowest limit as $x = 10^{-4}$, the PDFs are extracted from CTEQ6 dataset for momentum transfer $Q^2 = 10^4$ GeV$^2$ [48]. The cross sections are not susceptible to the choice of $Q^2$ [1,7]. In Fig. 2, the scattering cross sections corresponding to CC and NC processes are illustrated. At $x \to 0$, the PDFs are divergent. In this region since there is no experimental constraint, consideration of different behaviour of PDFs result in different scattering cross sections [49]. However, UHE neutrinos ($10^{10}$ eV) can probe upto $x \sim 10^{-4}$ at $M_{Pl,4+n} \sim 1$ TeV.

The NSI parameters used in this analysis are obtained from the global analysis of oscillation and COHERENT data considering CP conserving variables [12]. In COHERENT experiment the axial vector NSI couplings are neglected [50], hence in our analysis we consider them to be zero. Also the contribution of the strange quark is considered to be zero [50–52]. The values of NSI parameters used in our analysis are given in Table 1.

In the SM, only flavour diagonal NC processes $\nu_\alpha N \to \nu_\alpha X$ ($\alpha = e, \mu, \tau$) are allowed. These three processes have equal cross sections as the SM Wilson coefficients are independent of the incoming neutrino flavour. Unlike SM, NSI can allow NC flavour off-diagonal processes $\nu_\alpha N \to \nu_\beta X$ ($\alpha \neq \beta$) in addition to flavour conserving interactions ($\alpha = \beta$). The cross sections for CC and NC processes are depicted in Fig. 2. For CC process, the contribution to the cross section comes only from the SM interaction. From the top panel of the plot, it can be observed that the NSI provides marginal enhancement in the scattering cross section. Consequently these processes have marginal contribution to the total number of events generated by the incoming neutrinos. From the bottom panel of Fig. 2, it can be seen that the flavour non-diagonal or FCNC processes have much lower scattering cross section in comparison to the the flavour conserving interactions. As a result, the FCNC processes have insignificant contributions to the total number of events.

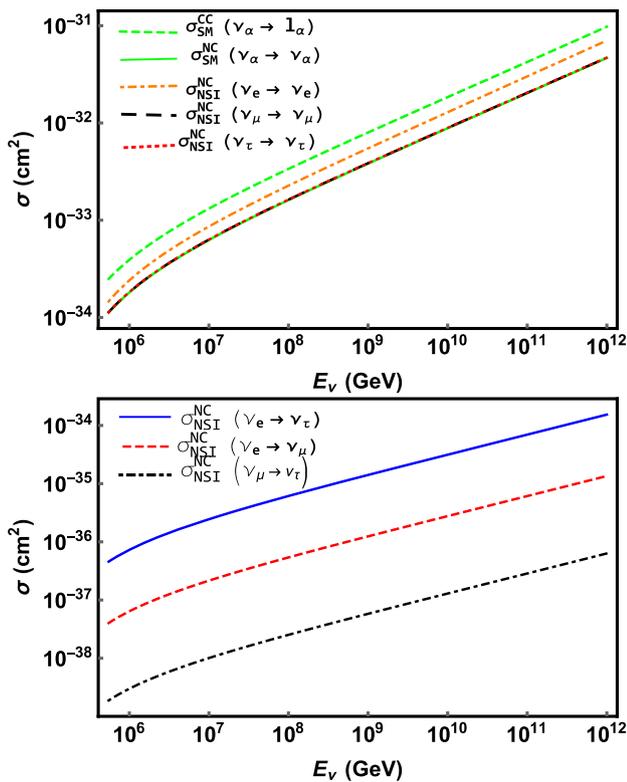

**Fig. 2** Variation of scattering cross sections ($\sigma$) in CC and NC processes. Top panel: For CC and NC flavour diagonal processes (NC includes both SM and NSI); bottom panel: For flavour non-diagonal processes in case of NC interactions (with NSI)

**Table 1** Values of NSI parameters from COHERENT experiment in $2\sigma$ range including the information of both the time and energy [12]. The contributions of $\varepsilon_{\alpha\beta}^{s,V}$ is zero, as well as $\varepsilon_{\alpha\beta}^{f,A}$ for all the three quarks ($f \in u, d, s$)

| Bounds on $\varepsilon_{\alpha\beta}^{u(d),V}$ | | | |
|---|---|---|---|
| $\varepsilon_{ee}^{u,V}$ | [0.043, 0.384] | $\varepsilon_{ee}^{d,V}$ | [0.036, 0.354] |
| $\varepsilon_{e\mu}^{u,V}$ | [−0.055, 0.027] | $\varepsilon_{e\mu}^{d,V}$ | [−0.052, 0.024] |
| $\varepsilon_{e\tau}^{u,V}$ | [−0.14, 0.09] | $\varepsilon_{e\tau}^{d,V}$ | [−0.106, 0.082] |
| $\varepsilon_{\mu\mu}^{u,V}$ | [−0.05, 0.062] | $\varepsilon_{\mu\mu}^{d,V}$ | [−0.046, 0.057] |
| $\varepsilon_{\mu\tau}^{u,V}$ | [−0.006, 0.006] | $\varepsilon_{\mu\tau}^{d,V}$ | [−0.005, 0.005] |
| $\varepsilon_{\tau\tau}^{u,V}$ | [−0.05, 0.065] | $\varepsilon_{\tau\tau}^{d,V}$ | [−0.046, 0.059] |

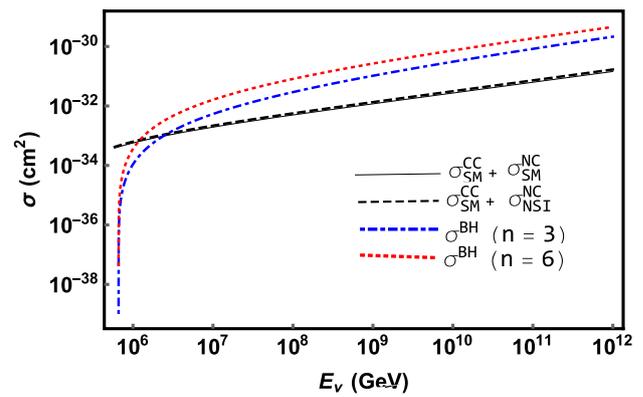

**Fig. 3** Variation of total cross section (i.e. the sum of CC and NC interactions, NC includes both the cases of SM and NSI) along with the cross section of black hole formation

**Table 2** Number of shower events created in the process $\nu N \to lX$ ($l = e, \mu, \tau$) for both SM and NSI. $N_e$, $N_\mu$ and $N_\tau$ are the number of events created individually by $\nu_e$, $\nu_\mu$ and $\nu_\tau$, respectively

| SM | | | | NSI | | | |
|---|---|---|---|---|---|---|---|
| $N_e$ | $N_\mu$ | $N_\tau$ | $N_{tot}$ | $N_e$ | $N_\mu$ | $N_\tau$ | $N_{tot}$ |
| 1.68 | 1.68 | 1.68 | 5.04 | 1.85 | 1.64 | 1.65 | 5.15 |

The cross section of BH formation ($\sigma^{BH}$) is illustrated in Fig. 3 for $n = 3$ and 6 extra dimensions at modified Planck scale $M_{Pl,4+n} = 1$ TeV. The total cross sections for neutrino-nucleon scattering processes for both CC and NC interactions in the presence of SM and NSI interaction are also shown in the figure for comparison. It can be observed that at UHE regime, the BH production cross section via $\sigma(\nu N \to BH \to lX)$ process is at least two orders of magnitude higher than $\sigma(\nu N \to lX)$. It should also be noted that the cross section of BH formation is independent of short distance physics.

The number of shower events generated by the process $\nu N \to lX$ are estimated from Eq. (11) and the comparative results for SM and NSI are illustrated in Table 2 in the context of IceCube experiment which operates over a broad range of energy and capable of detecting UHE neutrinos by optical signal detection processes. The results are obtained by considering the detector area $A \sim 1$ km$^2$ over a time period of $T = 1$ year assuming WB flux. For simplicity, the zenith angle dependence of the column density $X$ is ignored which provides a factor of two from the integral over the angle. From Table 2 it can be seen that the increment in the number of shower events due to NSI is inconsequential.

New physics can affect the number of BH events through $P_{surv}^{BH,i}$ via $\sigma^{NC}(E_{\nu i})$ term in Eq. (14). In Fig. 4, we represent the variation of $P_{surv}^{BH,i}$ with energy which shows that NSI is unable to provide any observable deviation from the SM prediction in the entire energy range. It is therefore evident

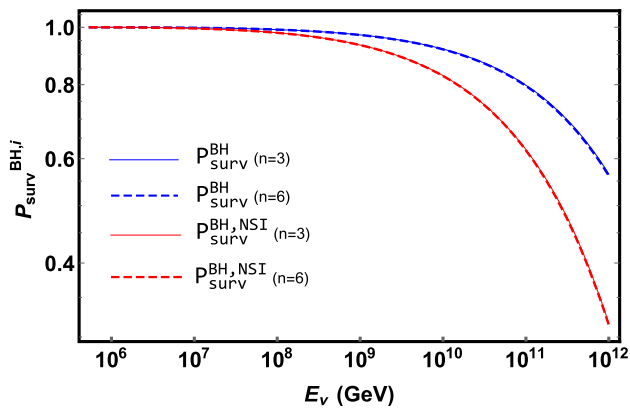

**Fig. 4** Variation of BH survival probability, $P_{surv}^{BH,i}$, with energy $E_{\nu_i}$ of a particular neutrino $\nu_i$ for the cases, $n = 3$ and $6$

that the effect of NSI on the number of BH events will also be insignificant.

## 4 Conclusion

We study the effects of NSI on the events generated by the scattering of UHE neutrinos with nucleons in the Earth's atmosphere. There events can be generated with or without the formation of microscopic BHs. We find that NSI can only provide marginal increase in the number of shower events produced in the absence of BH production. We also find that BH survival probability does not change significantly in presence of NSI. This implies that the number of events produced through BH production would remain nearly unaltered.

**Data Availability Statement** This manuscript has no associated data or the data will not be deposited. [Authors' comment: This is a theoretical study and no experimental data has been used.]